\documentclass[]{pasj02} 
\usepackage[switch,mathlines]{lineno} 

\jyear{2024}
\Received{}
\Accepted{}

\begin{document} 

\title{The unusual spectrum of the X-ray transient source XRISM J174610.8$-$290021 near the Galactic center}

\author{
 Anje \textsc{Yoshimoto}\altaffilmark{1}\altemailmark\email{yaa\_yoshimoto@cc.nara-wu.ac.jp},\, 
 Shigeo \textsc{Yamauchi}\altaffilmark{2},\,\,
 Masayoshi \textsc{Nobukawa}\altaffilmark{3},\,\,\orcid{0000-0003-1130-5363}
 Hideki \textsc{Uchiyama}\altaffilmark{4},\,\,\orcid{0000-0003-4580-4021}
 Kumiko K. \textsc{Nobukawa}\altaffilmark{5},\,\,\orcid{0000-0002-0726-7862}
 Yuma \textsc{Aoki}\altaffilmark{5},\,\,
 Manabu \textsc{Ishida}\altaffilmark{6},\,\,
 Yoshiaki \textsc{Kanemaru}\altaffilmark{6},\,\,\orcid{0000-0002-4541-1044}
 Megumi \textsc{Shidatsu}\altaffilmark{7},\,\,
 Takayuki \textsc{Hayashi}\altaffilmark{8,9,10},\,\,
 Yoshitomo \textsc{Maeda}\altaffilmark{6},\,\,\orcid{0000-0002-9099-5755}
 Hironori \textsc{Matsumoto}\altaffilmark{11,12},\,\,
 Yohko \textsc{Tsuboi}\altaffilmark{13},\,\,
 Hiromasa \textsc{Suzuki}\altaffilmark{14},\,\,\orcid{0000-0002-8152-6172}
 Hiroshi \textsc{Nakajima}\altaffilmark{15},\,\,\orcid{0000-0001-6988-3938}
 Q. Daniel \textsc{Wang}\altaffilmark{16},\,\,\orcid{0000-0002-9279-4041}
 Satoshi \textsc{Eguchi}\altaffilmark{17},\,\,\orcid{0000-0003-2814-9336}
Tomokage \textsc{Yoneyama}\altaffilmark{13},\,\,\orcid{0000-0002-2683-6856}
Tadayasu \textsc{Dotani}\altaffilmark{6, 18},\,\,
Ehud \textsc{Behar}\altaffilmark{19},\,\,\orcid{0000-0001-9735-4873}
Yukikatsu \textsc{Terada}\altaffilmark{6, 20},\,\,\orcid{0000-0002-2359-1857}
Nari \textsc{Suzuki}\altaffilmark{1},\,\,
  and Marina \textsc{Yoshimoto}\altaffilmark{7}\,\,
}

\altaffiltext{1}{Graduate School of Humanities and Sciences, Nara Women's University, Kitauoyanishimachi, Nara, Nara 630-8506, Japan}
\altaffiltext{2}{Faculty of Science, Nara Women's University, Kitauoyanishimachi, Nara, Nara 630-8506, Japan}
\altaffiltext{3}{Faculty of Education, Nara University of Education, Takabatake-cho, Nara 630-8528, Japan}
\altaffiltext{4}{Faculty of Education, Shizuoka University, 836 Ohya, Suruga-Ku, Shizuoka 422-8529, Japan}
\altaffiltext{5}{Faculty of Science and Engineering, Kindai University, 3-4-1, Kowakae, Higashi-Osaka 577-8502, Japan}
\altaffiltext{6}{Institute of Space and Astronautical Science (ISAS), Japan Aerospace Exploration Agency (JAXA), Sagamihara, Kanagawa 252-5210, Japan}
\altaffiltext{7}{Department of Physics, Ehime University, 2-5, Bunkyocho, Matsuyama, Ehime 790-8577, Japan}
\altaffiltext{8}{Center for Research and Exploration in Space Science and Technology (CRESST II), Greenbelt, MD 20771, USA}
\altaffiltext{9}{Department of Physics, University of Maryland, Baltimore County, 1000 Hilltop Circle, Baltimore, MD 21250, USA}
\altaffiltext{10}{NASA's Goddard Space Flight Center, X-ray Astrophysics Division, Greenbelt, MD 20771, USA}
\altaffiltext{11}{Department of Earth and Space Science, Graduate School of Science, Osaka University, 1-1, Machikaneyama-cho, Toyonaka, Osaka 560-0043, Japan}
\altaffiltext{12}{The Forefront Research Center, Graduate School of Science, Osaka University, 1-1, Machikaneyama-cho, Toyonaka, Osaka 560-0043, Japan}
\altaffiltext{13}{Department of Physics, Faculty of Science and Engineering, Chuo University, 1-13-27, Kasuga, Bunkyo-ku, Tokyo 112-8551, Japan}
\altaffiltext{14}{Faculty of Engineering, University of Miyazaki, Miyazaki 889-2192, Japan}
\altaffiltext{15}{College of Science and Engineering, Kanto Gakuinn University, 1-50-1 Mutsuura Higashi, Kanazawa-ku, Yokohama, Kanagawa 236-8501, Japan}
\altaffiltext{16}{Department of Astronomy, University of Massachusetts Amherst, 710 North PleasantStreet Amherst, MA 01003, USA}
\altaffiltext{17}{Faculty of Economics, Department of Economics, Kumamoto Gakuen University, 2-5-1, Oe, Chuo-ku, Kumamoto 862-8680, Japan}
\altaffiltext{18}{Graduate Institute for Advanced Studies, School of Physical Sciences, SOKENDAI, 3-1-1, Yoshino-dai, Chuou-Ku, Sagamihara, Kanagawa 252-5210, Japan}
\altaffiltext{19}{Department of Physics, Technion, Technion City, Haifa 32000, Israel}
\altaffiltext{20}{Graduate School of Science and Engineering, Saitama University, 255 Shimo-Okubo, Sakura, Saitama, Saitama 338-8570, Japan}




\KeyWords{Galaxy: center  ---  X-rays: stars  ---  X-rays: binaries  ---  stars: low-mass }  

\maketitle

\begin{abstract}
The Galactic center region was observed with the XRISM X-ray observatory during the performance verification phase in 2024 and a point-like X-ray source was detected with the X-ray imager Xtend at a position of $(\alpha, \delta)_{\rm J2000.0}=(\timeform{17h46m10.8s}, \timeform{-29D00'21''})$, which is thus named XRISM J174610.8$-$290021. This source was bright in February to March and showed time variations in count rate by more than one order of magnitude in one week. The 2--10~keV X-ray luminosity was $\sim\,10^{35}~\rm{erg~s^{-1}}$ for the assumed distance of 8 kpc. However, after six months, it was below the detection limit. We found a hint of periodicity of 1537~s from timing analysis. The XRISM/Xtend spectrum has emission lines from helium-like iron (Fe\,\emissiontype{XXV}-He$\alpha$) at 6.7~keV and hydrogen-like iron (Fe\,\emissiontype{XXVI}-Ly$\alpha$) at 6.97~keV; their intensity ratio is unusual with the latter being four times stronger than the former. If the emission is of thermal origin, the ionization temperature estimated from the iron-line intensity ratio is $\sim$\,30~keV, which is inconsistent with the electron temperature estimated from the thermal bremsstrahlung, $\sim$\,7~keV. Spectral models of magnetic cataclysmic variables, which are often seen in the Galactic center in this luminosity range, are found to fail to reproduce the obtained spectrum. By contrast, we found that the spectrum is well reproduced with the models of low-mass X-ray binaries containing a neutron star plus two narrow Gaussian lines. We consider that the source is intrinsically bright reaching $10^{37}~\rm{erg~s^{-1}}$, but is blocked from direct view due to a high inclination and only the scattered emission is visible. The photoionized plasma above the accretion disk with an ionization parameter of  $\sim\,10^{5}$ may explain the unusual iron line ratio. We further discuss the potential contribution of point sources of the type of XRISM J174610.8$-$290021 to the diffuse Galactic center X-ray emission. 
\end{abstract}


\section{Introduction}
One of the characteristics of the Galactic center (GC) region is the presence of the Galactic center X-ray emission (GCXE).  Transient sources in the GC region may be the constituent of the GCXE, and its nature should be understood in relation with the GCXE. 
The GCXE is unresolved X-ray emission with a luminosity of $(0.8$--$2.3)\,\times\,10^{37}~\rm{erg~s^{-1}}$ in the GC region (e.g., \cite{Koyama18} and references therein). 
The most distinctive feature in the GCXE spectrum is the three prominent emission lines at 6--7 keV, which are K lines from neutral (Fe\,\emissiontype{I}-K$\alpha$), helium-like (Fe\,\emissiontype{XXV}-He$\alpha$), and hydrogen-like (Fe\,\emissiontype{XXVI}-Ly$\alpha$) iron at 6.4, 6.7, and 6.97 keV, respectively \citep{Koyama07}. The latter two emission lines originate from high-temperature plasma with a temperature of $\sim$\,7~keV \citep{Uchiyama13}.

A part of the GCXE should be attributed to individual point-like X-ray sources, often referred to as X-ray active stars (XASs). 
Among many surveys conducted by various X-ray observatories in the past, Chandra surveys stood out in resolving and detecting XASs with its deep surveys of a $17'\,\times\,17'$ field around Sgr A$^{\star}$ (e.g., \cite{Muno03a}; \cite{Muno03b}; \cite{Muno04}) and a wider-field survey of $\timeform{2D}\,\times\,\timeform{0D.8}$ (e.g., \cite{Wang02}; \cite{Muno06}); they detected $\sim$\,8,000 XASs in total. 
\citet{Muno06} found from the Chandra GC catalog for $\timeform{2D}\,\times\,\timeform{0D.8}$ that these Chandra-detected XASs consisted of five groups: cataclysmic variables (CVs),  Wolf-Rayet and O (WR/O) stars, young isolated pulsars, low-mass X-ray binaries (LMXBs), and high-mass X-ray binaries (HMXBs). 
Among the five groups of XASs, CVs are the dominant ($\sim\,$90\%) population in the Chandra catalog, and the X-ray luminosities of the CVs in the catalog are $L_{X}\,\lesssim\,10^{33.5}~\rm{erg~s^{-1}}$. \citet{Revnivtsev06} and \citet{Yuasa12} considered that most of them are magnetic CVs (mCVs). 
A mCV is a binary comprising a Roche Lobe-filling low-mass star and a magnetized white dwarf. 
The post-shock plasma of the accretion flow formed at the magnetic poles of the white dwarf, emits optically thin thermal X-rays, whose continuum spectra are described by bremsstrahlung with a temperature of 10-40 keV (\cite{Ezuka99}; \cite{Ishida91}).

Low-mass X-ray binaries (LMXBs) are present in the GC, although in smaller numbers than CVs. The maximum X-ray luminosity of LMXBs can reach $10^{39}~\rm{erg~s^{-1}}$ \citep{Muno06}.  
A LMXB contains a black hole or a neutron star, together with a low-mass star, and the population of the latter (NS-LMXBs), each of which comprises a Roche Lobe-filling low-mass star and a neutron star, outnumbers that of the former. 
NS-LMXBs have two X-ray spectral states: hard state and soft state. 
Typically, the X-ray spectrum in the former state shows a power-law-like shape and can be represented by soft thermal and Comptonized components, and that in the latter state is reproduced by two optically thick components of a disk blackbody and a blackbody originating from the neutron star (e.g., \cite{Mitsuda84}; \cite{Mitsuda89}; \cite{Sakurai14}).

\citet{Wang02} first pointed out the point source origin of the GCXE.
About 40\% of the GCXE has been resolved into XASs with the threshold X-ray luminosity of $\sim\,10^{31}~\rm{erg~s^{-1}}$ (e.g., \cite{Revnivtsev07}). 
The remaining $\sim\,$60\% of the emission is not explained by the known populations of the CVs and active binaries from the comparison of the line equivalent width and geometrical scale of the iron lines in the GCXE (\cite{Nobukawa16}; \cite{Yamauchi16}). 
These studies suggest that additional components are needed to explain the GCXE and a candidate is a collection of undiscovered supernova remnants. 
Even now, the bulk of the GCXE remains unresolved and its origin continues to be discussed. 
A better understanding of individual objects in the GC can help to resolve this decades-long puzzle.

The XRISM (X-Ray Imaging and Spectroscopy Mission) X-ray observatory observed the GC during the performance verification phase in 2024 \citep{Tashiro25}. 
The observation serendipitously detected a point-like X-ray emission near the GC. This X-ray source has a peculiar spectral structure. 
In this work, we report the characteristics of the source and discuss its classification and its implication for understanding the GCXE.

The present paper is organized as follows. 
The observations and data reduction are described in section \ref{sec:2}. We report the analysis and results in section \ref{sec:3}. 
Specifically, section \ref{ssec:31}  presents the X-ray peak detection and counterpart search, section~\ref{ssec:32} describes the detected time variability,  and \ref{ssec:33} presents phenomenological and physical model fitting and also the phase-dependent spectral analysis. 
We discuss the classification and counterpart of the point source in section \ref{sec:4} and conclude in section \ref{sec:5}. 
In this paper, the distance of the GC is assumed to be $8~\rm{kpc}$, and its\ redshift, to be 0.0 for simplicity.

\begin{figure*}
 \begin{center}
  \includegraphics[width=15cm]{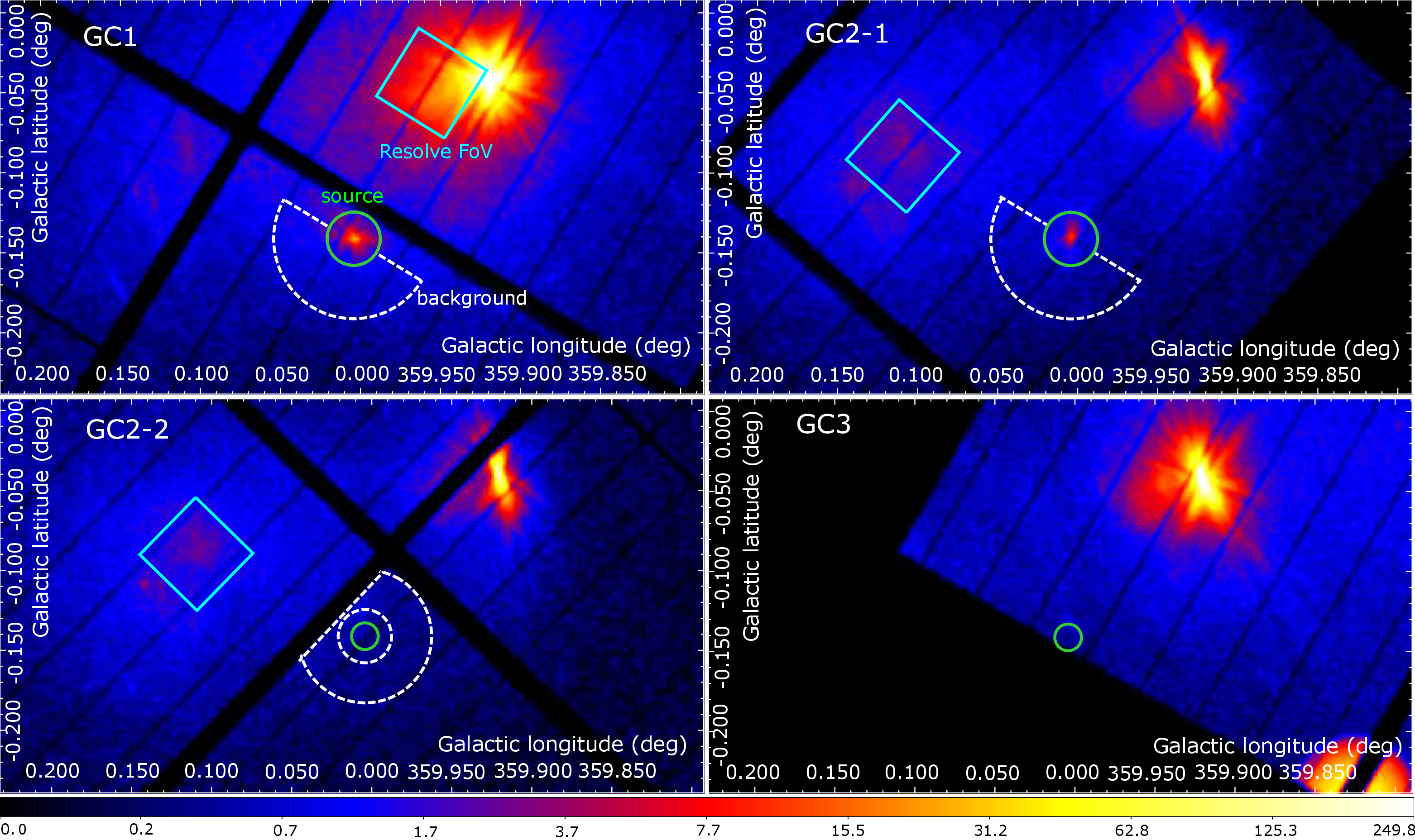}
 \end{center}
\caption{ Xtend images of the GC  in the energy range of 2--10~keV.
 {The color bar shows surface brightness  in $\rm{counts~s^{-1}~pixel^{-1}}$ on a logarithmic scale ($1~{\rm pixel}\,=\,1.8~{\rm arcsec}$). Upper-left, upper-right, lower-left, and lower-right panels show those of the observations from 2024 February 26 (ObsID: 300044010, GC1),  2024 February 29 (ObsID: 300045010, GC2-1),  2024 August 23 (ObsID: 300045020, GC2-2), and 2024 August 28 (ObsID: 300046010, GC3), respectively (table~\ref{tab:1}). A point-like source was detected in the GC1 and GC2-1 observations, but the source faded away six months later. The green circles show the source regions with  radii of $1'$ for GC1 and GC2-1 data and $30''$ for GC2-2 and GC3 data, and the  incomplete annuli of  white  dashed lines show the background regions. Cyan squares delineate the Resolve FoVs ($3'\times3'$)  for reference. The calibration source at lower right corner was not masked in the GC3 image. Alt text: Four X-ray images.
}
}\label{fig:1}
\end{figure*}

\begin{table}
  \tbl{ XRISM observation log of the GC. }{%
  \begin{tabular}{lccc}
      \hline 
      Data & ObsID &  Period & Exposure [ks]\footnotemark[$*$]  \\ 
      \hline
      GC1 & 300044010 & 2024 February 26--29 & 107.9 \\
      GC2-1 & 300045010 & 2024 February 29--March 2 & 63.3\\
      GC2-2 & 300045020 & 2024 August 23--25 & 53.8 \\
      GC3 & 300046010 & 2024 August 28--31 & 111.8 \\
      \hline
    \end{tabular}}\label{tab:1}
\begin{tabnote}
\footnotemark[$*$] Exposure time after data screening.  \\ 
\end{tabnote}
\end{table}

\section{Observation and data reduction}\label{sec:2}
The XRISM observations of the GC were carried out in four occasions from 2024 February 26 (ObsID: 300044010) and 29 (ObsID: 300045010), and 2024 August 23 (ObsID: 300045020) and 28 (ObsID: 300046010), each lasting for a few days. 
Here, we refer to these four datasets as GC1, GC2--1, GC2--2, and GC3, respectively. 
The GC2-1 and GC2-2 have the same aim point but different roll angles, whereas the others have different aim points. Also, GC1 and GC2-1 are continuously observed. The basic parameters of the four observations are summarized in table~\ref{tab:1}. 
XRISM has an X-ray microcalorimeter, Resolve \citep{Ishisaki22}, and an X-ray CCD camera, Xtend (\cite{Mori22}; \cite{Noda25}). 
Xtend has 4 CCDs and covers a $38'\,\times\,38'$ field of view (FoV) for the energy range of 0.4--12~keV. 
The energy resolution of Xtend is $\sim$\,170~eV in full width at half-maximum (FWHM) at 6~keV \citep{Uchida25}.


In this paper, only Xtend data were used for the subsequent analysis as is described in section \ref{ssec:31}.
The Xtend was operated in the full-window mode for all the GC data.
In the data reduction, periods of the Earth eclipse and sunlit Earth’s limb, and South Atlantic Anomaly (SAA) passages were excluded. 
We performed standard data processing, and the exposures shown in the table~\ref{tab:1} remained for the point-like sources.

We reduced these data with HEAsoft version 6.34 (HEASARC 2014) and calibration database version 9 (v241115). 
We generated a redistribution matrix file (RMF) with {\tt xtdrmf} task using the cleaned event file and CALDB. Also, we made a auxiliary response file (ARF) by the {\tt xaarfgen} assuming a point-like source at the point of the source (see section \ref{ssec:31}).
In our spectral analysis, we use the software package XSPEC version 12.14.1 \citep{Arnaud96}  with the Chi-square statistic.
In addition, XSTAR version 2.58e, which is a computer program that calculates the physical conditions, spectra of ionized and nearly neutral gases, and warmabs version 2.49d, a package for using the results calculated with XSTAR for spectral model fitting, were used for the analysis.
We use an energy range of 2--10~keV for our analysis. The errors quoted for spectral analyses hereafter are at a 90\% confidence level.






\section{Analysis and results}\label{sec:3}

\begin{table}
  \tbl{ The count rates, fluxes, and luminosities of XRISM J174610.8$-$290021. }{%
  \begin{tabular}{lccc}
      \hline 
      Data & Count rate & Flux\footnotemark[$*$]  & Luminosity\footnotemark[$\dag$] \\ 
      \quad & [$10^{-2} \rm{count\,s^{-1}}$] & [$10^{-11}\rm{erg\,s^{-1}\,cm^{-2}}$] & [$10^{34}\rm{erg\,s^{-1}}$]\\
      \hline
      GC1 & $12.8$ & 1.8 & 13 \\
      GC2-1 & $5.5$ & 1.2 & 9.0\\
      GC2-2 & $<\,$0.07 & $<\,$0.017 & $<\,$0.13 \\
      \hline
    \end{tabular}}\label{tab:2-1}
\begin{tabnote}
\footnotemark[$*$] The energy range is 2--10~keV. Assuming the model in section\ref{ssec:332}.\\
\footnotemark[$\dag$] The energy range is 2--10~keV. Assuming the distance of $8~\rm{kpc}$.\\
\end{tabnote}
\end{table}

\subsection{X-ray image}\label{ssec:31}
\begin{table*}
  \tbl{ Cataloged X-ray sources in SIMBAD located within a radius of $20$~arcsec from the XRISM X-ray peak  position of XRISM J174610.8$-$290021. }{%
  \begin{tabular}{lcccc}
      \hline 
      Source name & Galactic longitude [deg] & Galactic latitude [deg] &  Offset [arcsec] & Remarks\\
      \hline
  CXOU J174610.8$-$290019 & $0.00458$ & $-0.14070$ & 1.2 & Type I X-ray burst\footnotemark[$*$]\\
  4XMM J174610.7$-$290020 & $0.00432$ & $-0.14039$ & 1.6 & -- \\
  CXO J174610.7$-$290019 & $0.00456$ & $-0.14024$ & 2.4 & -- \\
  SWIFT J174610.4$-$290018 & $0.00433$ & $-0.13911$ & 6.3 & Low luminosity outburst\footnotemark[$\dag$] \\ 
  CXOGCS J174610.26$-$290015.1 & $0.00465$ & $-0.13831$ & 9.2 & Extended source\footnotemark[$\ddag$] \\
  CXOGCS J174610.8$-$290010 & $0.00024$ & $-0.14334$ & 17.1 & Point source\footnotemark[$\S$] \\
  2CXO J174610.8$-$290038 & $0.00499$ & $-0.13614$ & 17.1 & --\\
      \hline
    \end{tabular}}\label{tab:2}
\begin{tabnote}
\footnotemark[$*$] Reference is \citet{Pastor-Marazuela20}.\\
\footnotemark[$\dag$] Reference is \citet{Reynolds24}.\\
\footnotemark[$\ddag$] Reference is \citet{Wang06}.\\
\footnotemark[$\S$] Reference is \citet{Muno03a}.\\
\end{tabnote}
\end{table*}

Figure~\ref{fig:1} shows the XRISM/Xtend images of the GC in each observation epoch after binnig with $1~{\rm pixel}\,=\,1.8~{\rm arcsec}$.
We detected a point-like source in the first two observations (GC1 and GC2-1; from 2024 February to March) within the Xtend FoV but did not detect it in the later observations GC2-2 or GC3 (2024 August). We estimate the count rates of GC1 and GC2 in the 2--10 keV band to be $(12.8\,\pm\,0.1)\,\times\,10^{-2}$ and $(5.5\,\pm\,0.1)\,\times\,10^{-2}~\rm{counts~s^{-1}}$, respectively, by subtracting the GCXE and non-X-ray background simultaneously taken from the background region (see section \ref{ssec:32}). 
Then, the source was at least two orders of magnitude fainter in the later GC2-2 and GC3 observations;  the background-subtracted count rate in GC2-2 was $<\,7\times\,10^{-4}~\rm{counts~s^{-1}}$ at a 90\% confidence level. The source was not in the FoV of Resolve in any of the observations. In this work, we analyzed the Xtend data only. 
Since the data from GC3 was at the edge of the FoV and similar in the brightness limit to GC2-2, only GC2-2 was included here. We list the count rates, fluxes, luminosities of the source estimated from the GC1, GC2-1, and GC2-2 data in table~\ref{tab:2-1} (using the  model in section \ref{ssec:332}).

We estimated the position of this source using the GC1 image (2--10 keV),  which has the best statistics among our four data sets. 
The image was smoothed with a Gaussian with a radius of $6.0~{\rm pixel}$ and $\sigma\,=\,3.0~\rm{pixel}$ ($1~{\rm pixel}\,=\,1.8~{\rm arcsec}$), using the software SAOImageDS9 version 8.3 (Smithsonian Astrophysical Observatory, 2000). 
The X-ray peak position for the source on the smoothed image was found to be $(\alpha, \delta)_{\rm J2000.0}=(\timeform{17h46m10.8s}, \timeform{-29D00'21''})$, or $(l, b)=(\timeform{0.0043D}, \timeform{-0.1408D})$ in the Galactic coordinates. 
The statistical error of the source position estimated from the Lorentzian fitting of a projection is 10~arcsec at $1\,\sigma$ error. 
The pointing determination accuracy of Xtend is smaller than 20 arcsec \citep{Kanemaru24}. 
From the X-ray peak coordinate, we designated it XRISM J174610.8$-$290021 accordingly.

The SIMBAD catalog \citep{Wenger00} lists seven X-ray sources within a radius of 20 arcsec from the XRISM X-ray peak (table~\ref{tab:2}) but no infrared sources. 
Among the listed sources, CXOU J174610.8$-$290019 is the closest X-ray source with a separation angle of 1.2 arcsec from the XRISM X-ray peak. CXOGCS J174610.26$-$290015.1 is least likely to be the counterpart, considering that it was a extended source, whereas the XRISM source was time-variable (discribed in the next subsection).
ATel $\#$16481 \citep{Reynolds24} reported that the X-ray telescope Swift detected SWIFT J174610.4$-$290018  on 2024 February 22, which is close to the first observation (GC1) date of the GC with XRISM. 
Given the relatively poor detection threshold of Swift, SWIFT J174610.4$-$290018 is a strong contender for the counterpart of XRISM J174610.8$-$290021.
Note that the Swift source may well be one of the Chandra sources in the error circle (table~\ref{tab:2}).

\subsection{Time variability}\label{ssec:32}
Figure~\ref{fig:1} displays the source and background regions of each observation for our timing and spectral analyses. 
For GC1 and GC2-1 (when the source was bright and detected), we chose the circle region with a radius $1'$ from the X-ray peak position for the source region and a concentric semi-circular region with a radius $1'\,<\,r\,<\,3'$, excluding the source region, at the far side from Sgr A$^{\star}$ for the background, minimize the spatial variation of the GCXE and also the contamination from the transient source AX J1745.6$-$2901 that was situated close to Sgr A$^{\star}$ and was very bright during GC1. 
For GC2-2 (when the source was undetected), we selected the circle region with a radius of $30''$ from the X-ray peak for the source region and an annular region with radii of $1'$--$2.5'$ arcsec centered at the source position but excluding a small region in and beyond the CCD chip gap (figure~\ref{fig:1}).  
The background contributions for each data remained the same, even with changed in the background regions.

\begin{figure}
 \begin{center}
  \includegraphics[width=6cm,angle=270]{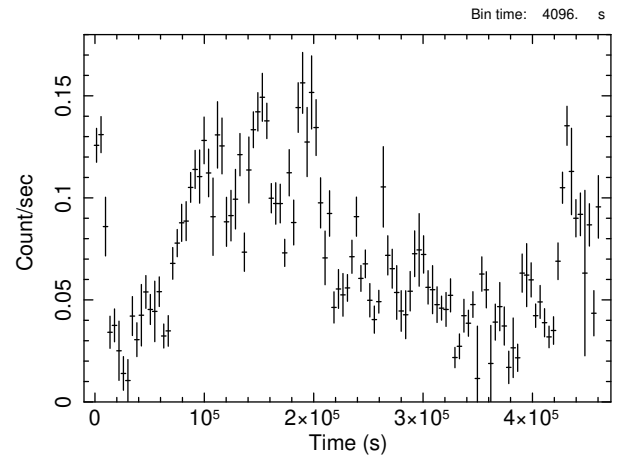}
 \end{center}
\caption{Light curve of observations GC1 and\ GC2-1 (2--10 keV) with a bin size of 4096~s. 
 {The background is subtracted from the source light curve. Alt text: A graph showing the intensity variation.} 
}\label{fig:2}
\end{figure}

Figure~\ref{fig:2} shows the 2--10 keV background-subtracted light curve in GC1 and GC2-1, where the source was significantly detected. 
Hereafter, we subtracted the background by using the light curve of the background region. 
The source clearly shows time variability in the count rates by more than one order of magnitude in one week.
Fast variation on time scale of $\sim 10$~minutes is also seen, constraining the source size of $\lesssim 10^{13}$~cm.
Hence, the source is likely a Galactic compact stellar object.
We note that neither X-ray burst nor periodic dip structure was found.

\begin{figure*}
 \begin{center}
    \includegraphics[width=5cm,angle=270]{image/efsearch_sub_gc121_2-10_1535s_1s_241230_250129.eps}
    \includegraphics[width=5cm,angle=270]{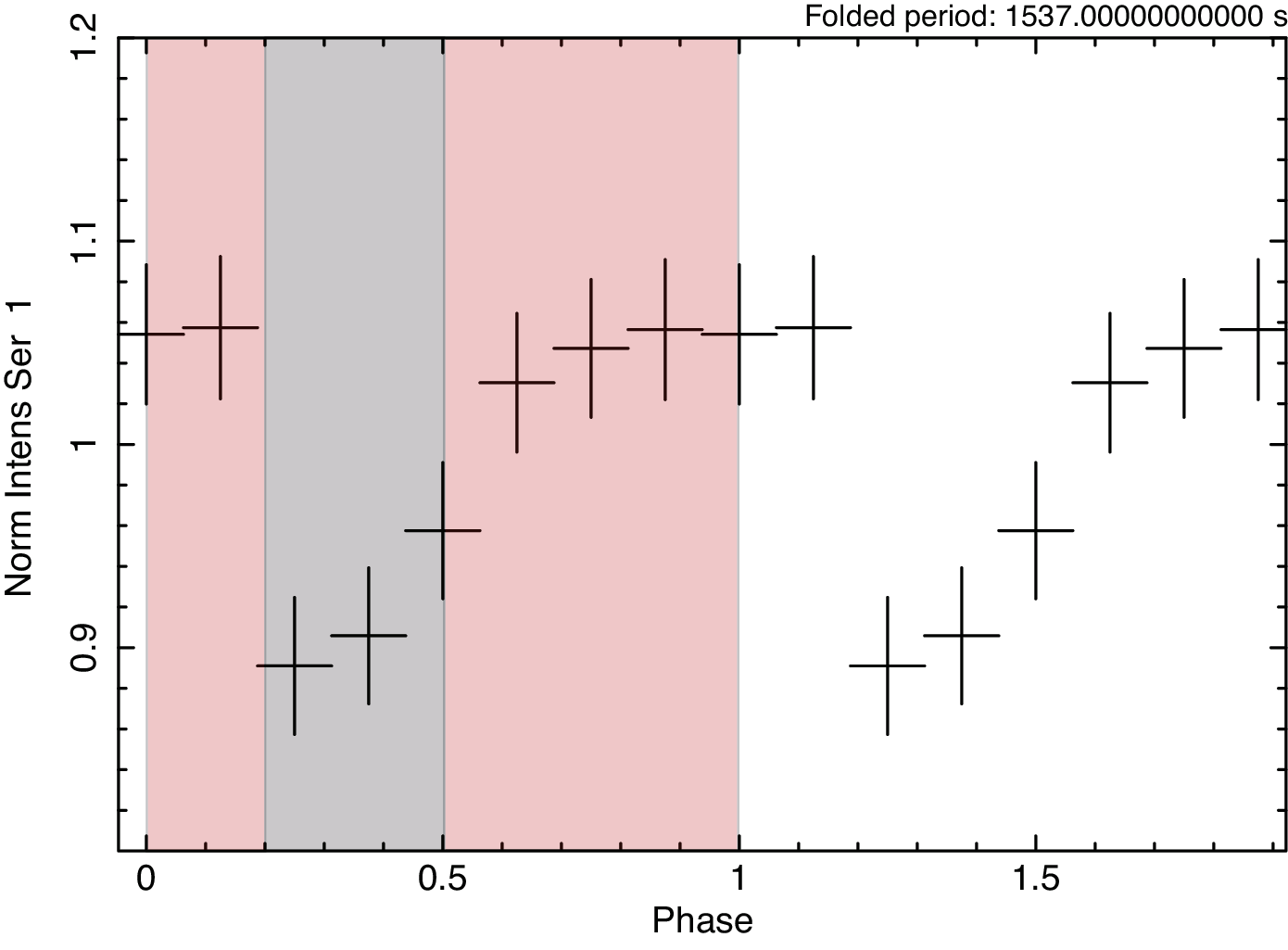}
 \end{center}
\caption{Searching results for periodicities of XRISM J174610.8$-$290021 in the 2--10 keV band.
 {
 Left panel: The result of epoch-folding search for a maximum chi-square for periods around 1537~s. The search is performed for 128 periods with a step of  1.0 s. Right panel: Folded light curve for the best-estimated period of 1537~s for two complete cycles. The red and gray shadows indicate the boundaries of the phases (high phase for 0--0.2 \& 0.5--1.0, and low phase for 0.2--0.5) used for the phase-dependent spectral analysis (see section~\ref{ssec:333}). Alt text: Two line graphs.
} 
}\label{fig:3}
\end{figure*}

We searched for potential periodicity in the light curve of this source, using the package Xronos version 6.0. 
We first derived the power spectrum and found two peaks at $\sim\,5.0\times10^{-4}~\rm{Hz}$ and $\sim\,6.3\times10^{-4}~\rm{Hz}$, which correspond to 2000~s and 1600~s, respectively. 
Then, we searched for the period with a maximum chi-square, using the epoch folding technique at around the candidate periods. 
We calculated the chi-square values for 128 periods with a step of $\Delta\,=\,4.0~\rm{s}$ for each candidate period. 
Then, we did the same calculation again with a finer resolution of $\Delta\,=\,1.0~\rm{s}$ and determined the most likely periods to be  1924~s and 1537~s.  
We notice that the former is $1\,/\,3.0$ of the orbital period of XRISM, suggesting a possibility that the period is not intrinsic to the source. 
To evaluate it, we did the same analysis for candidate periods of the orbital period divided by integers and found some signals. 
This fact confirms that the obtained former candidate period of 1924~s is not intrinsic to the source. 
The latter, 1537~s, is $\sim\,1\,/\,3.7$ times the satellite orbital period and is clearly unrelated to it. 
The result of epoch-folding search for periods around 1537~s shown in the left panel of figure~\ref{fig:3} indicated that the chi-square peak is to be $\sim\,$30 ($d.o.f.\,=\,7$). 
As we have 128 trial periods, a chance probability to get this value of chi-square from statistical fluctuation would be 1.2\%. 
Thus, the significance of the peak is marginal, and we should interpret it as just a hint of periodicity.
We folded the background-subtracted 2--10 keV light curve with the period 1537~s and show it in the right panel of figure~\ref{fig:3}.

\begin{figure}
 \begin{center}
  \includegraphics[width=6cm,angle=270]{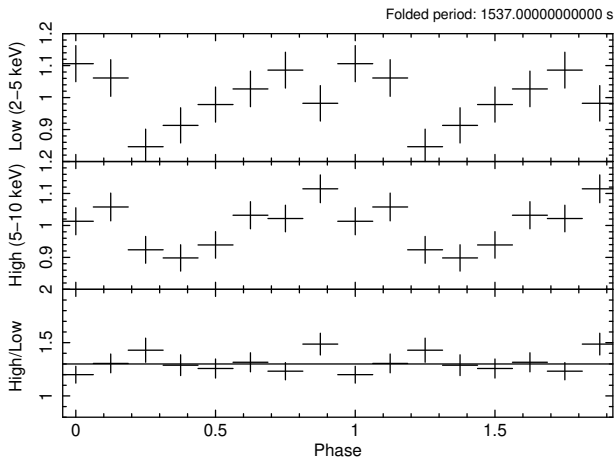}
 \end{center}
\caption{Folded light curves in the high (2--5~keV) and low (5--10~keV) energy bands and its hardness ratio for the period of 1537~s.
{
Alt text: Three line graphs.
}
}\label{fig:4}
\end{figure}

We reduced the energy-divided folded light curves for the period of 1537~s in the energy bands 2--5~keV and 5--10~keV and its hardness ratio (figure \ref{fig:4}). 
Fitting of the hardness-ratio light-curve with a constant model yielded an acceptable result of $\chi^2/d.o.f.\,=\,7.23/7$, indicating no significant variability in the hardness or spectral shape.

\subsection{X-ray spectrum}\label{ssec:33}
We made spectra of each data from the source and background regions for XRISM J174610.8$-$290021. 
Especially for the GC1 and GC2--1 data, we merged two spectra at counts and subtracted the merged background spectrum from the merged source spectrum. 
Figure~\ref{fig:5} shows the resultant background-subtracted spectrum. 
The background level is 30\% in 2--10 keV and 24\% in Fe line energy range (6.5--7.3 keV). 
We also confirmed that there was no variation in the subtracted spectrum even if we changed the selection of the background region.
The response files of the two datasets were merged with weights according to the photon statistics.\\
  
\subsubsection{Phenomenological model fit}\label{ssec:331}

\begin{table*}
  \tbl{Best-fit parameters with the phenomenological models. }{%
  \begin{tabular}{llcccccc}
      \hline 
        Component & Parameter & Bremsstrahlung & Power-law & CIE plasma ({\tt{APEC}})\\ 
      \quad & \quad & plus two Gaussian lines & plus two Gaussian lines &\quad \\
      \hline
       Absorption & $N_{\rm{H}}$~[$10^{22}$~cm$^{-2}$] & 20.9$^{+1.3}_{-1.2}$ & 23.9$^{+1.7}_{-1.6}$ & 18.5$^{+1.1}_{-0.9}$\\
       \hline
       Continuum & $kT_{e}$~[keV] & 6.8$^{+1.0}_{-0.8}$ & -- & 9.6$^{+1.0}_{-1.2}$\\
       \quad & Abundance~[solar] & -- & -- & 0.30$^{+0.12}_{-0.10}$ \\
       \quad & Photon\ Index & -- & 2.38$^{+0.13}_{-0.14}$ & -- \\
        \hline
       He-like Fe line\footnotemark[$*$] & Center Energy [keV] & 6.70$^{+0.09}_{-0.09}$ & 6.70$^{+0.08}_{-0.08}$ &  -- \\
       \quad & Intensity\footnotemark[$\ddag$]  & 0.7$^{+0.4}_{-0.4}$ & 0.7$^{+0.4}_{-0.4}$ & -- \\
       \quad & Equivalent Width [$10$~eV] & 5$^{+3}_{-3}$ & 5$^{+3}_{-3}$ & -- \\
       \hline
       H-like Fe line\footnotemark[$\dag$] & Center Energy [keV] & 6.99$^{+0.02}_{-0.02}$ & 6.99$^{+0.02}_{-0.02}$ & -- \\
        \quad & Intensity\footnotemark[$\ddag$]  & 2.3$^{+0.4}_{-0.4}$ & 2.4$^{+0.4}_{-0.4}$ & -- \\
        \quad & Equivalent Width [$10^{2}$~eV] & 1.9$^{+0.4}_{-0.4}$ & 2.0$^{+0.4}_{-0.4}$ & -- \\
        \hline
        $\chi^2\,/\,d.o.f.$ & \quad &  166.93\,/\,151 & 179.38\,/\,151 & 228.34\,/\,154\\ 
      \hline
    \end{tabular}}\label{tab:3}
\begin{tabnote}
\footnotemark[$*$] Line width is fixed to 35~\rm{eV}.  \\ 
\footnotemark[$\dag$] Line width is fixed to 0~\rm{eV}. \\
\footnotemark[$\ddag$] In  $10^{-5}~\rm{photons~cm^{-2}~s^{-1}}$. \\
\end{tabnote}
\end{table*}

\begin{figure}
 \begin{center}
  \includegraphics[width=8.5cm,angle=0]{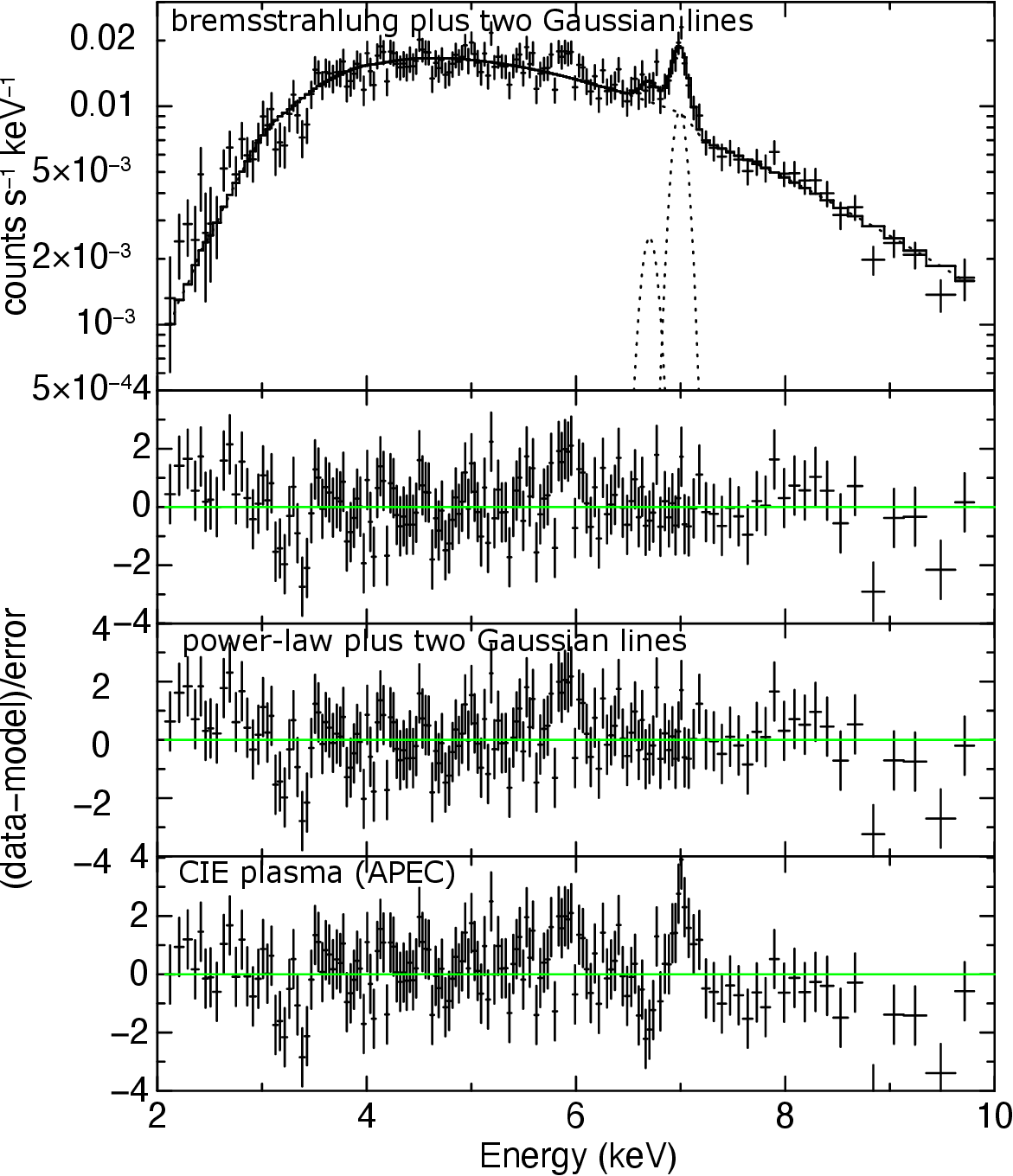}
 \end{center}
\caption{ Results of the spectral fitting with the phenomenological models.
 {In the top panel, crosses are the data points, and the histogram is the best-fit model composed of thermal bremsstrahlung with photoelectric absorption plus two narrow Gaussian lines. The dotted lines show the individual model components. The bottom three panels show the residuals of the best-fit results for the (upper panel) above-mentioned model, (middle panel) a power-law with photoelectric absorption plus two narrow Gaussian lines, and (lower panel) CIE plasma with photoelectric absorption. Alt text: A line graph showing the best-fitting result of three phenomenological models to the data.}
}\label{fig:5}
\end{figure}

The most remarkable feature of the spectrum of XRISM J174610 is a strong emission line at around $\sim$\,7~keV, the most dominant component of which was identified as the Fe\,\emissiontype{XXVI}-Ly$\alpha$ line. 
In addition, the spectrum shows a continuum extending to at least 10~keV and strong absorption.
These facts imply that the source has a high-temperature plasma with photoelectric absorption. 
We first applied to the spectrum a phenomenological model consisting of thermal bremsstrahlung ({\tt{bremss}} in XSPEC) with photoelectric absorption plus two narrow Gaussian lines for Fe\,\emissiontype{XXVI}-Ly$\alpha$ and neighboring Fe\,\emissiontype{XXV}-He$\alpha$ and obtained a reasonably good fit ($\chi^2/d.o.f.\,=\,166.93/151$). 
Here and hereafter, the sigma of the two Gaussian lines for Fe\,\emissiontype{XXV}-He$\alpha$ and Fe\,\emissiontype{XXVI}-Ly$\alpha$ were fixed to 35~eV and 0 eV, respectively. The best-fit result is shown in figure~\ref{fig:5}, while the best-fit parameters are listed in table~\ref{tab:3}. 
The large interstellar column density $N_{\rm H}\gtrsim 10^{23}$~cm$^{-2}$ suggests that the source is located as far as the GC region.
The Fe\,\emissiontype{XXVI}-Ly$\alpha$ line center energy is determined to be $6.99\,\pm\,0.02$~keV. 
Comparing the value with the theoretical one of $6.97$~keV, no significant redshift is detected. 
We hereafter assume that the source is located at the same distance as to the GC (8.0~kpc).
We measured the 2--10~keV flux to be $1.6\,\times\,10^{-11}~\rm{erg~s^{-1}~cm^{-2}}$, which corresponds to an unabsorbed 2--10~keV luminosity of $1.2\,\times\,10^{35}~\rm{erg~s^{-1}}$.
The best-fit ratio of Fe\,\emissiontype{XXVI}-Ly$\alpha$\,/\,Fe\,\emissiontype{XXV}-He$\alpha$ was $4^{+5}_{-2}$;  the intensity of Fe\,\emissiontype{XXVI}-Ly$\alpha$  was much stronger than that of Fe\,\emissiontype{XXV}-He$\alpha$. 
The ionization temperature estimated from the line intensity ratio was $\sim$\,30~keV, much higher than the electron temperature $\sim$\,7~keV estimated from the bremsstrahlung.

The structure of line-like emission is also seen around 5.9~keV in the spectrum. 
We carefully checked possible contamination of the calibration source ($^{55}$Fe) illuminating a corner of the CCD chips, but the source location is far away from the illuminated corner and such possibility is unlikely. In fact, we see no similar structure in the background spectra.
We checked the significance by adding a Gaussian line with the sigma of 0~eV to the above model and performing the F-test from two chi-square values. 
This model approximated the spectrum with $\chi^2/d.o.f.\,=\,152.75/149$ and the center energy of the line-like structure was estimated to be $5.91\,\pm\,0.04$~keV. 
The center energy is about the same as the Cr\,\emissiontype{XXIV}-Ly$\alpha$ line energy.
From the result of the model fitting, we calculated the F-statistic as an indicator and its probability to be 6.920 and 0.00134, respectively. 
The F-test indicated that the significance level of the 5.9~keV line is $\sim\,3\,\sigma$. 

An application of a power-law function ({\tt powerlaw}) for the continuum instead of bremsstrahlung approximated the spectrum with $\chi^2/d.o.f.\,=\,179.38/151$. The result of this model fitting is acceptable within 2 sigma, but the hard energy band of the spectrum has residuals rather than the model fitting with bremsstrahlung for the continuum (see the residuals in figure~\ref{fig:5}).
The best-fit parameters with the power-law function are listed in table~\ref{tab:3}. The 2--10~keV flux according to this best-fit model was  $1.9\,\times\,10^{-11}~\rm{erg~s^{-1}~cm^{-2}}$.

Next, we fitted the spectrum with a single collisional ionization equilibrium (CIE) plasma model ({\tt{APEC}}\ in XSPEC). The residuals of the best-fit model are shown in figure \ref{fig:5}, while the best-fit parameters are listed in table \ref{tab:3}. 
The result shows large residuals at $\sim$\,7~keV, implying a failure in reproducing the observed prominent Fe\,\emissiontype{XXVI}-Ly$\alpha$ line, with $\chi^2/d.o.f.\,=\,228.34/154$. 
This result is consistent with the above-mentioned fact of a large discrepancy between the ion and electron temperatures.

Using the phenomenological model, we estimated the upper limit of the 2--10~keV flux in observation GC2-2  to be $<\,1.7\,\times\,10^{-13}~\rm{erg~s^{-1}~cm^{-2}}$, which is smaller than the average flux during observations GC1 and GC2-1 by about two orders of magnitude. 

\subsubsection{Physical model fit}\label{ssec:332}

\begin{figure*}
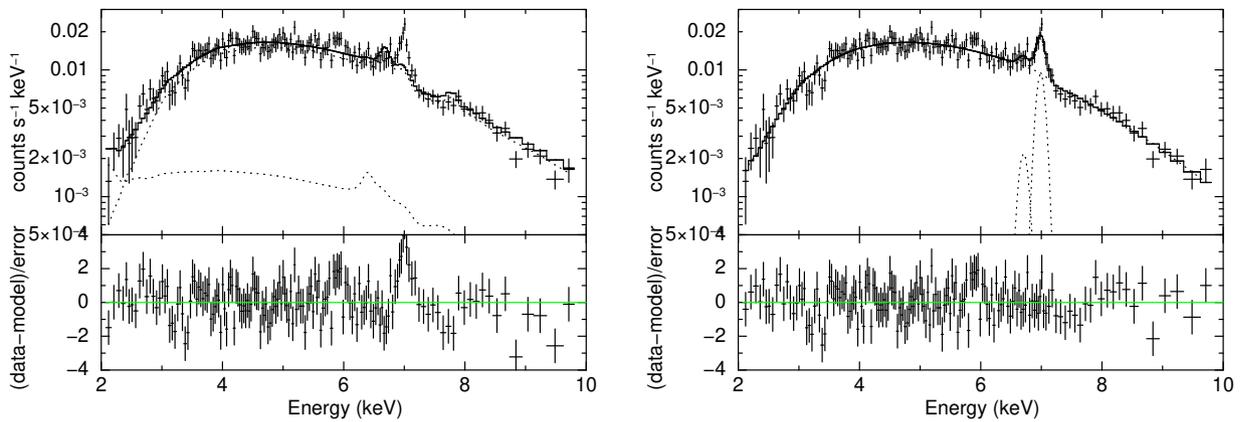

 \begin{center}
  \includegraphics[width=5.4cm,angle=270]{image/specfit_hayashi_gc121_photo_2-10_250422.eps}
    \includegraphics[width=5.4cm,angle=270]{image/specfit_compPSgaga_gc121_photo_2-10_250422.eps}
 \end{center}
\caption{ Xtend spectrum (in crosses) of  XRISM J174610.8$-$290021 with the best-fit model and fitting residuals (in the lower panels). Histograms in the upper panels show the best-fit models, and dotted lines do their individual components. The applied models are
 {(Left panel)  a mCV model \citep{Hayashi14a} and   (Right panel) a NS-LMXB with the low luminosity of $L_{X}\,\lesssim\,10^{35}~\rm{erg~s^{-1}}$} model (Comptonized spectral model assuming blackbody for the seed photon source with photoelectric absorption plus two Gaussian lines). Alt text: Two line graphs showing the best-fitting result of models for a magnetic cataclysmic variable and a neutron star low mass X-ray binary to the data.
}\label{fig:6}
\end{figure*}

\begin{table*}
  \tbl{Best-fit parameters  with the physical-model fitting of the observed XRISM J174610.8$-$290021 spectrum}{%
  \begin{tabular}{llcccccc}
      \hline 
        Component & Parameter &Disk blackbody and blackbody & Comptonized spectral model\footnotemark[$\dag\dag$] \\ 
       \quad & \quad & plus two Gaussian line & plus two Gaussian line\\
      \hline
       Absorption & $N_{\rm{H}}$~[$10^{22}$~cm$^{-2}$] & 13.7$^{+3.6}_{-1.1}$ & 13.0$^{+1.6}_{-1.0}$ \\
       \hline
       Continuum & $kT_{in}$~[keV]\footnotemark[$\ddag$] & 1.03 & --\\
       \quad & $kT_{bb}$~[keV]\footnotemark[$\S$] & 1.60$^{+0.12}_{-0.06}$ & 1.59$^{+0.04}_{-0.18}$\\
       \quad & $kT_{e}$~[keV]\footnotemark[$\|$] & -- & 100.0\\
       \quad &$R_{bb}$~[km] \footnotemark[$\sharp$] & 0.33$^{+0.03}_{-0.05}$ & 0.34$^{+0.11}_{-0.02}$\\
       \quad & $\tau$ & -- & $\,<\,0.5$\\
        \hline
       He-like Fe line\footnotemark[$*$] & Center Energy [keV] & 6.71$^{+0.10}_{-0.10}$ & 6.71$^{+0.09}_{-0.09}$\\
       \quad & Intensity\footnotemark[$**$] & 0.5$^{+0.4}_{-0.4}$ & 0.5$^{+0.4}_{-0.4}$\\
       \quad & Equivalent Width [$10$~eV] & 4$^{+3}_{-3}$ & 4$^{+3}_{-3}$\\
       \hline
       H-like Fe line\footnotemark[$\dag$] & Center Energy [keV] & 6.99$^{+0.02}_{-0.012}$ & 6.99$^{+0.02}_{-0.02}$\\
        \quad & Intensity\footnotemark[$**$] & 2.1$^{+0.4}_{-0.4}$ & 2.1$^{+0.4}_{-0.4}$\\
        \quad & Equivalent Width [$10^{2}$~eV] & 1.8$^{+0.4}_{-0.4}$ & 1.9$^{+0.4}_{-0.4}$\\
        \hline
        $\chi^{2}\,/\,d.o.f.$ & \quad & 149.81\,/\,150 & 149.89\,/\,150\\ 
      \hline
    \end{tabular}}\label{tab:4}
\begin{tabnote}
\footnotemark[$*$] Line width is fixed to 35~\rm{eV}.  \\ 
\footnotemark[$\dag$] Line width is fixed to 0~\rm{eV}. \\
\footnotemark[$\ddag$] Temperature at the inner disk radius. This parameter was calculated using $T_{bb}:T_{in}=0.760:0.488$.\\
\footnotemark[$\S$] Blackbody temperature.     \\
\footnotemark[$\|$] Electron temperature in the corona. This parameter is fixed in the fitting.\\
\footnotemark[$\sharp$] $R_{bb}$ is the source radius assuming the distance to the source to be 8~kpc.\\
\footnotemark[$**$] In $10^{-5}~\rm{photons~cm^{-2}~s^{-1}}$.\\ 
\footnotemark[$\dag\dag$] Seed photon source is blackbody from the neutron star.\\
\end{tabnote}
\end{table*}

A typical point source that has an X-ray luminosity of $L_{X}\,\sim\,10^{35}~\rm{erg~s^{-1}}$, a high-temperature continuum, and iron K-lines would be a mCV. 
Whereas a single CIE model cannot reproduce the obtained spectrum of XRISM J174610.8$-$290021 (section \ref{ssec:331}), a typical mCV spectrum consists of a multi-temperature plasma model and thus may possibly reproduce it well.  
We chose the {\tt{ACRAD}} model (\cite{Hayashi14a}; \cite{Hayashi14b}) as the representative spectral model of the mCV, which considers a post-shock accretion column of intermediate polars and has four parameters: the white dwarf mass, specific accretion rate,  metal abundance, and normalization. 
The application of the {\tt{ACRAD}} model to the spectrum  failed to reproduce the spectrum  with\ $\chi^2/d.o.f.\,=\,239.60/152$. 
The best-fit result is shown in the left panel of figure~\ref{fig:6}. 
In the fitting, we determine an upper limit of the white-dwarf mass to $<\,0.5\,M_{\odot}$. 
Nevertheless, the fitting  failed to simultaneously reproduce the iron-line intensities and continuum.

Another candidate point source class for XRISM J174610.8$-$290021 that has a hard X-ray continuum is the NS-LMXB. 
Some of the known NS-LMXB  have X-ray luminosities of $L_{X}\,\sim\,10^{35}~\rm{erg~s^{-1}}$ (e.g., Aquila~X-1 in the hard state, \cite{Sakurai12}), and some of them have strong iron emission lines (e.g., accretion disk corona, \cite{Iaria13}).
We applied to the obtained spectrum of XRISM J174610.8$-$290021 the typical NS-LMXB model, which consists of a disk blackbody ({\tt{diskbb}}) and blackbody ({\tt{bbodyrad}}) undergoing photoelectric absorption.
Also, we added two Gaussian models for the typical model to represent the iron lines.
In the model-fitting, the black body temperature ($T_{bb}$) was linked to the temperature at the inner disk radius ($T_{in}$) for \texttt{diskbb} with the relation  $T_{bb}\,:\,T_{in}\,=\,0.760\,:\,0.488$ \citep{Frank02} because it is impossible to directly determine the temperature $T_{in}$ from the observed spectrum due to strong absorption.
The model was found to well reproduce the spectrum with $\chi^2/d.o.f.\,=\,149.81/150$.
The best-fit parameters are listed in table \ref{tab:4}. Notably,  the flux was determined to be $1.1\,\times\,10^{-11}~\rm{erg~s^{-1}~cm^{-2}}$ (2--10~keV), which is converted to the 2--10~keV X-ray luminosity of $8.1\,\times\,10^{34}~\rm{erg~s^{-1}}$, and the ratio of Fe\,\emissiontype{XXVI}-Ly$\alpha$/Fe\,\emissiontype{XXV}-He$\alpha$, to be $4^{+8}_{-2}$.   

According to \citet{Sakurai14}, we apply a Comptonized spectral model ({\tt compPS}) without the disk reflection signals which represent a NS-LMXB to the spectrum continuum instead of disk blackbody plus blackbody.
Here, we assumed that electrons scatter some of the seed photons emitted from the neutron star in the corona at a temperature of $kT_{e}=100~\rm{keV}$ and seed photon source of blackbody. 
We found that the model very well approximated the data with $\chi^2/d.o.f.\,=\,149.89/150$. 
The best-fit result is shown in the right panel of figure \ref{fig:6}, while the best-fit parameters are listed in table \ref{tab:4}. 
An upper limit of 0.5 was obtained for optical depth.
The flux was estimated to be $1.0\,\times\,10^{-11}~\rm{erg~s^{-1}~cm^{-2}}$ (2--10~keV), which is converted to  the 2--10~keV luminosity of $7.9\,\times\,10^{34}~\rm{erg~s^{-1}}$. 
The line intensity ratio Fe\,\emissiontype{XXVI}-Ly$\alpha$\,/\,Fe\,\emissiontype{XXV}-He$\alpha$ was $4^{+7}_{-2}$. 


In the case of black hole binary, the multi-color disk component with an inner disk temperature of $\lesssim\,$1.5~keV (e.g., Cyg X-1, \cite{Dotani97}; XTE J1550$-$564, \cite{Kubota04}; V4641 Sgr, \cite{Shaw22}) is dominant in high\,/\,soft state, whereas the spectrum in low\,/\,hard state shows a hard spectrum due to the Compton scattering.
When we applied the {\tt diskbb} model to the source spectrum, the inner disk temperature was estimated to be $kT_{in}\,\sim\,2.4$~keV, inconsistent with the expected value in high\,/\,soft state.
In the case of the low/hard state, the temperature of the seed photon is higher than the typical inner disk temperature value of $<\,1$~keV (e.g., GRO J1655$-$40, \cite{Takahashi08}; GX 339$-$4, \cite{Shidatsu11}).
Also, the typical spectrum of high-mass X-ray binaries, which is explained by power-law with the photon index of $\Gamma\,\sim\,1.0-2.0$ and cutoff at 10--20~keV \citep{Lewin95}, is also inconsistent with our result. Hence, we make no further mention of the possibility of the black-hole binary and the high-mass X-ray binary for the XRISM J174610.8$-$290021.

\subsubsection{Phase-dependent spectral analysis}\label{ssec:333}
\begin{figure}
 \begin{center}
  \includegraphics[width=5.6cm,angle=270]{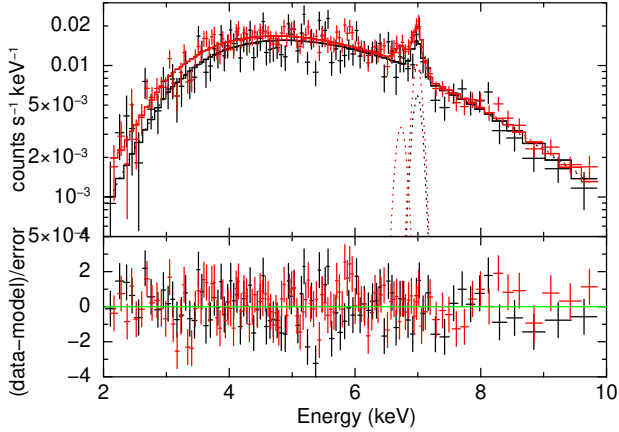}
 \end{center}
\caption{Phase-dependent spectra of  XRISM J174610.8$-$290021. 
 {The notations are the same as in figure~\ref{fig:6}. The model is a Comptonized spectral model assuming blackbody for the seed photon source with photoelectric absorption plus two narrow Gaussian lines (the same as in the right panel of figure~\ref{fig:6}). Black and red colors denote those in the low and high phases, respectively. Alt text: A line graph showing the best-fitting result of models for a neutron star low mass X-ray binary to the two data of different phase.} 
}\label{fig:7}
\end{figure}

We made the spectra of the two phase-intervals of high phase for 0.0--0.2\,\&\,0.5--1.0 and low phase for 0.2--0.5 (right panel in figure \ref{fig:3}) and simultaneously fitted them with {\tt compPS} model to investigate potential periodic variations of its hydrogen column density and Fe K$\alpha$ line ratio. 
Here, we linked all parameters except the hydrogen column density and line intensities. The best-fit results of each model fitting are shown in figure \ref{fig:7}, while the best-fit parameters are listed in table \ref{tab:5}. 
The best-fit hydrogen column densities for the high and low phases were 
$N_{\rm{H}}\,=\,12.7^{+1.7}_{-1.2}\,\times\,10^{22}~\rm{cm}^{-2}$ and $N_{\rm{H}}\,=\,15.1^{+2.1}_{-1.6}\,\times\,10^{22}~\rm{cm}^{-2}$, respectively; No significant difference was observed.
We calculated the ratios Fe\,\emissiontype{XXVI}-Ly$\alpha$\,/\,Fe\,\emissiontype{XXV}-He$\alpha$ in the respective phases to be $2.7^{+2.8}_{-1.0}$ and $>\,2.5$, which were consistent. 
In addition, the line-like structure around 5.9~keV seems to be seen only in the high-phase spectrum with significance $3\,\sigma$ from the F-test. 

The hardness ratio shown in figure \ref{fig:4} was constant and no variations in $N_{\rm{H}}$ of each phase were observed.
Also, we compared spectra in the higher and the lower intensity phases and found no significant difference in the spectral shape.

\begin{table}
  \tbl{Best-fit parameters of phase-dependent spectra with a Comptonized spectral model assuming blackbody for the seed photon source with two Gaussian lines.}{%
  \scalebox{0.95}{
  \begin{tabular}{llccc}
      \hline 
       Model & Parameter & High phase & Low phase\\ 
      \hline
       Absorption  & $N_{\rm{H}}$~[$10^{22}$~cm$^{-2}$] & 12.7$^{+1.7}_{-1.2}$ & 15.1$^{+2.1}_{-1.6}$\\
       \hline
       Continuum & $kT_{bb}$~[keV]\footnotemark[$\ddag$]\footnotemark[$\S$]  & 1.56$^{+0.07}_{-0.20}$ & = High phase \\
        \quad & $kT_{e}$~[keV]\footnotemark[$\ddag$]\footnotemark[$\|$] & 100.0 & = High phase\\
        \quad & $R_{bb}$~[km]\footnotemark[$\ddag$]\footnotemark[$\sharp$] & 0.35$^{+0.13}_{-0.03}$ & = High phase\\
        \quad & $\tau$\footnotemark[$\ddag$] & 0.10 (\,$<$\,0.65) & = High phase\\
        \hline
       He-like Fe line\footnotemark[$*$] & Center Energy [keV]\footnotemark[$\ddag$] & 6.73$^{+0.09}_{-0.07}$ & = High phase\\
       \quad & Intensity\footnotemark[$**$]  & 0.8$^{+0.4}_{-0.4}$ & 0.02 ($<\,$0.55)\\
       \quad & Equivalent Width [$10$~eV] & 7$^{+4}_{-4}$ & 0.1 ($<$\,5) \\
       \hline
       H-like Fe line\footnotemark[$\dag$] & Center Energy [keV]\footnotemark[$\ddag$]  & 6.99$^{+0.02}_{-0.02}$& = High phase\\
        \quad & Intensity\footnotemark[$**$] & 2.3$^{+0.5}_{-0.5}$ & 1.4$^{+0.7}_{-0.7}$\\
        \quad & Equivalent Width [$10^{2}$~eV] & 2.1$^{+0.4}_{-0.4}$ & 1.3$^{+0.6}_{-0.6}$\\
        \hline
        $\chi^2\,/\,d.o.f.$ & \quad & \multicolumn{2}{c}{234.59\,/\,198}\\
        \hline
    \end{tabular}}
    }\label{tab:5}
\begin{tabnote}
\footnotemark[$*$] Line width is fixed to 35~\rm{eV}.\\
\footnotemark[$\dag$] Line width is fixed to 0~\rm{eV}.\\
\footnotemark[$\ddag$] Link between two phase.\\
\footnotemark[$\S$] Blackbody temperature.  \\ 
\footnotemark[$\|$] Electron temperature in the corona. This parameter is fixed in the fitting.\\
\footnotemark[$\sharp$] $R_{bb}$ is the source radius assuming the distance to the source to be 8~kpc.\\ 
\footnotemark[$**$] In  $10^{-5}~\rm{photons~cm^{-2}~s^{-1}}$.\\
\end{tabnote}
\end{table}

\section{Discussion}\label{sec:4}
In this study, we found a strong absorption, thermal spectral continuum extending up to at least 10~keV, and peculiar iron-line intensity ratio from XRISM J174610.8$-$290021. 
In this section, we discuss the classification of this object and its counterpart from our spectral analysis results.

\subsection{Classification}\label{ssec:41}
\begin{figure*}
\begin{center}
 \includegraphics[width=14cm]{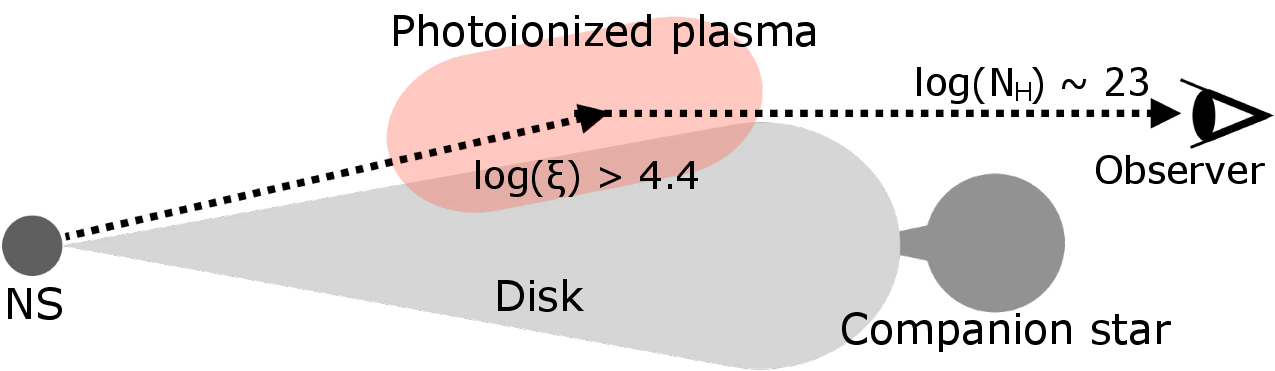}
 \end{center}
\caption{Proposed geometry of XRISM J174610.8$-$290021.
 {The system should be an NS-LMXB (see text). Dashed-line arrows show the path of X-rays to the observer. Alt text: A schematic view showing the position of the neutron star, photoionized plasma, and the observer.} 
}\label{fig:8}
\end{figure*}

We considered two candidate classes of X-ray sources in the spectral analysis (section~\ref{ssec:332}),  the mCV and NS-LMXB. 
Here, we evaluate their plausibility. The continuum of the typical mCV spectrum can be explained by thin-thermal bremsstrahlung with a temperature of 10--40~keV, and its spectrum also has iron lines of Fe\,\emissiontype{XXV}-He$\alpha$ and Fe\,\emissiontype{XXVI}-Ly$\alpha$ from the post-shock thermal plasma and Fe\,\emissiontype{I}-K$\alpha$ from cold matter irradiated by X-rays at the white dwarf surface \citep{Ishida91}. 
In this sense, the observed Xtend spectrum resembles that of a mCV. 
However, the Fe\,\emissiontype{I}-K$\alpha$ line lacks in the spectrum, and the result of our spectral model fitting with bremsstrahlung showed the electron temperature of $\sim$\,7~keV (section~\ref{ssec:332}), which is lower than the known typical electron temperatures of mCVs shown above. 
In addition, we found that the physical model of the mCV presented by \citet{Hayashi14a} cannot reproduce the continuum and thermal iron lines at the same time (section~\ref{ssec:332}). 
The X-ray luminosity of $\sim\,10^{35}~\rm{erg~s^{-1}}$ is higher than the mCV found before, either.
Hence, we consider that XRISM J174610.8$-$290021 is unlikely to be a mCV.

The typical spectrum of the other candidate class, the NS-LMXB, has multiple blackbody components from the accretion disk and a blackbody component from the neutron star.
A fitting with a typical NS-LMXB model explained the spectrum well. 
Furthermore, the Comptonized spectral model assuming blackbody for the seed photon source, which is known to reproduce well the spectra of NS-LMXBs, approximated the continuum shape too (section~\ref{ssec:332}).



We estimated the size of the emission region to be 0.33~km from the spectral analysis with two NS-LMXB models. 
It is much smaller than the typical size of the NS, and means that we see only a small fraction of the emission from the NS surface. 
We estimate the intrinsic luminosity by assuming that the blackbody emission originates from the whole surface of the NS.
Using the effective temperature converted from the color temperature of blackbody emission 1.6~keV based on \citet{Shimura95} and a typical NS size $\sim\,10$~km, the luminosity is estimated to be $10^{37}~\rm{erg\,s^{-1}}$. 
It is two orders of magnitude higher than the observed value, $10^{35}~\rm{erg\,s^{-1}}$.
Therefore, we consider that the NS is probably hidden in the accretion disk.


NS-LMXBs usually emit neither significant Fe\,\emissiontype{XXV}-He$\alpha$ nor Fe\,\emissiontype{XXVI}-Ly$\alpha$ lines, unlike our observed Xtend spectrum of XRISM J174610.8$-$290021, and this fact seems to contradict the hypothesis of the NS-LMXB classification. Given that this object likely has a very high X-ray luminosity of $L_{X}\,\sim\,10^{37}~\rm{erg~s^{-1}}$, a possible mechanism to explain the observed strong iron lines is the photoionization in the surrounding gas cloud that is irradiated with strong X-rays from the vicinity of the neutron star. 
Here, we evaluate this scenario. 
We note that the scenario must also explain the observed peculiar line-intensity ratio, i.e., the much stronger hydrogen-like iron line than the helium-like one. 

The $\xi$ parameter, one indicator of photoionization, is given by $\xi\,=\,L_{X}\,/\,n_{e}R^{2}$, where $R$ is the distance from the surface of the neutron star to the photoionized plasma and $n_{e}$ is the electron density of the photoionized plasma  (\cite{Tarter69}; \cite{Kallman04}).
As for the $\xi$ parameter, \citet{Kallman04} and \citet{Kallman19} showed $\log(\xi)\,>\,4.0$ when a Fe\,\emissiontype{XXVI}-Ly$\alpha$ emission line appears. 
To confirm the possibility of the photoionized scenario for the structure of Fe lines, we used a {\tt photemis} model in the X-ray spectral analysis software XSTAR/warmabs, which reproduce the photoionized emitters. 
We used the default population files, because the assumed ionizing spectrum, power-law of $\Gamma\,=\,2$, for them is close to that of XRISM J174610.8$-$290021.
We apply {\tt compPS} plus {\tt photemis} with photoelectric absorption model to the spectrum.
Here, we fixed all parameters of {\tt compPS} to the results of model fitting with {\tt compPS} plus two Gaussian lines, except the normalization (see section~\ref{ssec:332}).
Also, the abundance of Fe was fixed at 1.5 solar, and the rest were fixed at 1.0 solar.
We found that the Fe line structures of the spectrum were well reproduced by {\tt photemis} model with $\log(\xi)\,>\,4.4$.

A line-like feature was seen around the 5.9~keV in the spectrum. 
From the result of fitting the spectrum with bremsstrahlung adding a Gaussian line at 5.9~keV, the center energy was estimated to be $5.91\pm 0.04$~keV and significance was about $3\,\sigma$ according to the F-test (see section~\ref{ssec:331}). 
The residuals still remained in the result of the spectral analysis using {\tt photemis} set to solar abundance. 
Since the central energy corresponds to Cr\,\emissiontype{XXIV}-Ly$\alpha$, we tried to explain the residuals by Cr and found that 60 solar was needed, which is too large, and peculiar conditions must be considered.
Also, we see the 5.9~keV line-like structure only in the high-phase spectrum, which may be important to consider the origin.
Additional observations by XRISM/Resolve would be needed to see if this is true.

For the hint of periodicity estimated to be 1537~s, we compared it with the orbital period range of LMXBs from 0.19~hr to 398~hr \citep{Lewin95} and found that our result falls within the known range of the orbital period.
However, given that the orbital period is very close to the shortest one, we cannot exclude such possibility that the periodicity happened to be produced by the superposition of random time variations of the source. 
It is difficult to identify the mechanism of the periodicity from our marginal result.

On the basis of these evaluations, we propose that XRISM J174610.8$-$290021 is a NS-LMXB associated with a photoionized region. 
The apparent X-ray luminosity was considered lower than the typical NS-LMXBs as the source is blocked from the direct view probably by the thick accretion disk and only the scattered emission by the photoionized plasma (accretion disk corona) is observable. 
Hence, we suggest that the geometry of this transient source is as indicated in figure~\ref{fig:8}. 
From the difference between the X-ray luminosity expected from the blackbody temperature and obtained from the analysis, the inclination angle should be large, and XRISM observed the source from the edge of the accretion disk.
Some NS-LMXBs are accompanied by photoionized plasma and show absorption lines in the blackbody continuum from NS. 
This is because the photoionized plasma, or the absorber, is located in the line of sight of the NS.
In our case that the emission from the NS is not seen directly, absence of absorption lines may be naturally ecxplained.

\subsection{Counterparts}\label{ssec:42}
We compare the analysis results of this target with the two most probable counterparts (section~\ref{ssec:31}): CXOU J174610.8$-$290019 (spatially closest to the XRISM J174610.8$-$290021 peak position) and SWIFT J174610.4$-$290018 (closest observation date).

First, CXOU J174610.8$-$290019 was detected with Chandra \citep{Muno06}, and it was also later detected with XMM-Newton with a 2--10~keV flux of $1.2\,\times\,10^{-13}~\rm{erg~s^{-1}~cm^{-2}}$ \citep{Pastor-Marazuela20}. 
During our observations GC1 and GC2-1, the source was two orders of magnitude brighter than the previous study. 
XMM-Newton detected the Fe\,\emissiontype{XXV}-He$\alpha$ line but did not detect significant Fe\,\emissiontype{XXVI}-Ly$\alpha$ line in the spectrum. 
Also, the 100~s outburst was observed from CXOU J174610.8$-$290019 with XMM-Newton, but XRISM did not detected similar phenomenon during the present observations.
The behavior of CXOU J174610.8$-$290019 is different, but it would be the same object as XRISMJ174610.8$-$290021. 
XRISM may have observed it in a different state from those in the past observations.

Next, we discuss SWIFT J174610.4$-$290018. It showed an outburst lasting for 3--6 weeks, and Swift estimated with the 10-ks-exposure XRT observation during its outburst the hydrogen column density and the 2--10~keV flux to be $19.3_{-4.6}^{+5.3}\,\times \,10^{22}~\rm{cm^{-2}}$ and $1.8_{-0.6}^{+1.2}\,\times\,10^{-11}~\rm{erg~s^{-1}~cm^{-2}}$, respectively,  with the model of a power-law attenuated with a photoelectric absorption (ATel \#16642, \cite{Degnaar24}). 
These parameters are consistent with the best-fit parameters in this study estimated with the same model. Since SWIFT J174610.4$-$290018 is classified as a low-luminosity LMXB, we suggest that this is the same object as XRISM J174610.8$-$290021. 
Althoug it is stated that SWIFT J174610.4$-$290018 is a new transient (ATel \#16481, \cite{Reynolds24}), we still consider there is possibility  that SWIFT J174610.4$-$290018 and CXOU J174610.8$-$290019
are the same source, given that the latter's positional error is $\sim\,5.0$~arcsec and well encompasses the former's position.

\subsection{Contribution to the GCXE}\label{ssec:43}
We discuss the contribution to the GCXE of the population of point sources of the type of XRISM J174610.8$-$290021. 
The equivalent widths (EWs) of  Fe\,\emissiontype{XXV}-He$\alpha$ and  Fe\,\emissiontype{XXVI}-Ly$\alpha$ of the GCXE were derived to be $\sim500$ and $\sim200$~eV, respectively (\cite{Nobukawa16}), and those of this source are $(4\,\pm\,3)\,\times\,10~\rm{eV}$ and $(1.9\,\pm\,0.4)\,\times\,10^{2}~\rm{eV}$ (section~\ref{ssec:332}). 
Whereas the values of the latter are similar, those of the former are very different. 
Chandra observations suggest that CVs are the predominant population by $\sim 90$\% and LMXBs are minorities in the GC \citep{Muno06}.

XRISM J174610.8$-$290021 is intrinsically brighter with the X-ray luminosity of $1\,\times\,10^{38}~\rm{erg~s^{-1}}$, and thus accompanies photoionized plasma that emits Fe emission lines. The luminosity was estimated to be $\sim\,10^{35}~\rm{erg~s^{-1}}$ because we happened to observe the source from the edge on.
Since this situation is rare, we would not expect to find other similar objects in the GC.

Therefore, the contribution of point sources of the type of XRISM J174610.8$-$290021 to the GCXE should be insignificant.
If previously undiscovered types of sources with strong Fe emission lines ($\rm{EW}\,\sim\,100-500~\rm{eV}$) and low luminosity (especially below the detection limit of the Chandra observations) exist, those may contribute to the GCXE. 
We encourage the observation of iron emission line spectra with an instrument that has a wide FoV such as XRISM/Xtend.


\section{Conclusions}\label{sec:5}
We detected the X-ray transient source with peculiar Fe lines in the XRISM/Xtend GC observations in 2024 and identified it as a probable NS-LMXB with photoionized plasma.

From an X-ray image of the 2024 February data with XRISM, we determined the X-ray peak position of the transient source and designated it XRISM J174610.8$-$290021. 
The SIMBAD catalog lists seven candidate counterparts in its positional error circle. 
Among them, SWIFT J174610.4$-$290018, which was discovered shortly before our first XRISM observations, and CXOU J174610.8$-$290019, the position of which is close to the peak position of the XRISM source, maybe the same object. 

The source showed time variations by more than an order of magnitude, which clearly suggest that it is a point source. 
A hint of a period from the source was detected at 1537~s.

The spectrum shows a very strong line at 6.97 keV corresponding to Fe\,\emissiontype{XXVI}-Ly$\alpha$ and another line at 6.7 keV corresponding to Fe\,\emissiontype{XXV}-He$\alpha$. 
With the phenomenological model fitting, the spectrum is reasonably well approximated with the model of thermal bremsstrahlung with the electron temperature $\sim\,$7~keV plus two Gaussian lines for the above-mentioned lines.
The iron line intensity ratio is estimated to be $4^{+5}_{-2}$, which is peculiar. 
The ionization temperature determined from the intensity ratio is $\sim$\,30~keV, which is significantly different from the electron temperature.

Although its luminosity, time variability, and the presence of the bright iron K-lines suggest at first look that the source is an mCV, the physical model of the mCV \citep{Hayashi14a} is found not to well reproduce the continuum and iron line intensities simultaneously.
By contrast, a model of the disk blackbody plus blackbody, which represents the typical NS-LMXB, well reproduces the spectrum, and so does Comptonized spectral model assuming blackbody for the seed photon source, which represents the NS-LMXB. 
The size of the emission region was determined to be 0.33~km, which is much smaller than the typical NS size.
The result shows that only a small fraction of the emission from the NS surface was observed.
From the result of blackbody temperature $kT_{bb}\,\sim\,1.6~\rm{keV}$ and typical NS size, the X-ray luminosity is predicted to be $10^{37}~\rm{erg\,s^{-1}}$ considering the difference between color and effective temperature. This luminosity is inconsistent with the observed luminosity $10^{35}~\rm{erg\,s^{-1}}$.
We consider that the source is hidden from direct view by the accretion disk and only the scattered emission is visible. This is possible if we observed the source from the edge of the accretion disk.
Assuming the photoionization scenario for the strong Fe\,\emissiontype{XXVI}-Ly$\alpha$ emission, $\xi$ is estimated to be $\sim\,10^{5}$.

This type of source is not numerous in the GC and may not contribute to the GCXE. We expected other undiscovered peculiar sources with iron emission lines to be observed with a instrument which has wide FoV such as XRISM/Xtend.

\begin{ack}
This work was supported by JST SPRING, Grant Number JPMJSP2115.
This work was also supported by JSPS KAKENHI Grant Numbers
24K00677, 21K03615(MN), 
23H00151(KKN), 
23K22548(YM), 
23H00128(HM), 
24K17093(HS), 
24K17105(YK), 
and NASA grant 80NSSC23K0738(DQW). 
Furthermore, this work was supported by JSPS Core-to-Core Program, (grant number: JPJSCCA20220002).
\end{ack}









\begin{thebibliography}{}
\bibitem[Arnaud(1996)]{Arnaud96}
  Arnaud, K. A. 1996, Astronomical Data Analysis Software and Systems V, eds. Jacoby G. and Barnes J., 17, ASP Conf. Series volume 101
\bibitem[Ezuka and Ishida(1999)]{Ezuka99}
  Ezuka, H., \& Ishida, M. 1999, \apjs, 120, 277
\bibitem[Degenaar et al.(2024)]{Degnaar24}
  Degnaar, N., et al. 2024, ATel, 16642
\bibitem[Frank, King, and Raine(2002)]{Frank02}
  Frank, J., King, A., \& Raine, D. 2002, Accretion Power in Astrophysics (England: Cambridge Univ Pr)
\bibitem[Dotani et al.(1997)]{Dotani97}
  Dotani, T., et al. 1997, \apj, 485, 2, L87-L90
\bibitem[Hayashi and Ishida(2014a)]{Hayashi14a}
  Hayashi, T., \& Ishida, M. 2014a, \mnras, 438, 3, 2267
\bibitem[Hayashi and Ishida(2014b)]{Hayashi14b}
  Hayashi, T., \& Ishida, M. 2014b, \mnras, 441, 4, 3718
\bibitem[Iaria et al.(2013)]{Iaria13}
  Iaria, R., Di Salvo, T., D'A\`\i, A., Burderi, L., Mineo, T., Riggio, A., Papitto, A., \& Robba, N. R. 2013, \aap, 549, A33
\bibitem[Ishida(1991)]{Ishida91}
  Ishida, M. 1991, PhD thesis, University of Tokyo
\bibitem[Ishisaki et al.(2022)]{Ishisaki22}
  Ishisaki, Y., et al. 2022, \procspie, 12181, 121811S
\bibitem[Kallman et al.(2004)]{Kallman04}
  Kallman, T. R., Palmeri, P., Bautista, M. A., Mendoza, C., \& Krolik, J. H. 2004, \apjs, 155, 675
\bibitem[Kallman et al.(2019)]{Kallman19}
  Kallman, T. R., et al. 2019, \apj, 874, 51
\bibitem[Kanemaru et al.(2024)]{Kanemaru24}
  Kanemaru, Y. 2024, Proc. SPIE, 13093, 130935V
\bibitem[Koyama et al.(2007)]{Koyama07}
  Koyama, K., et al. 2007, \pasj, 59, S245
\bibitem[Koyama(2018)]{Koyama18}
  Koyama, K. 2018, \pasj, 70, 1
\bibitem[Kubota and Makishima(2004)]{Kubota04}
  Kubota, A., \& Makishima, K., 2004, \apj, 601, 1, 428-438
\bibitem[Lewin, van Paradijs, and van den Heuvel(1995)]{Lewin95}
  Lewin, W. H. G., van Paradijs, J., \& van den Heuvel, E. P. J. ed. 1995, X-ray Binaries (England: Cambridge Univ Pr), 126
\bibitem[Mori et al.(2022)]{Mori22}
  Mori, K., et al. 2022, \procspie, 12181, 121811T
\bibitem[Muno et al.(2003a)]{Muno03a}
  Muno, M. P., et al. 2003a, \apj, 589, 225
\bibitem[Muno et al.(2003b)]{Muno03b}
  Muno, M. P., Baganoff, F. K., Bautz, M. W., Brandt, W. N., Garmire, G. P., \& Ricker, G. R. 2003b, \apj, 599, 465
\bibitem[Muno et al.(2004)]{Muno04}
  Muno, M. P., et al. 2004, \apj, 613, 1179
\bibitem[Muno et al.(2006)]{Muno06}
  Muno, M. P., et al. 2006, \apjs, 165,173
\bibitem[Mitsuda et al.(1984)]{Mitsuda84}
  Mitsuda, K., et al. 1984, \pasj, 36, 741
\bibitem[Mitsuda et al.(1989)]{Mitsuda89}
  Mitsuda, K., Inoue, H., Nakamura, N., \& Tanaka, Y. 1989, \pasj, 41, 97
\bibitem[Nobukawa et al.(2016)]{Nobukawa16}
  Nobukawa, M., Uchiyama, H., Nobukawa, K. K., Yamauchi, S., \& Koyama, K. 2016, \apj, 833, 268
\bibitem[Noda et al.(2025)]{Noda25}
  Noda, H., et al. 2025, \pasj accepted, arXiv:2502.08030
\bibitem[Pastor-Marazuela et al.(2020)]{Pastor-Marazuela20}
  Pastor-Marazuela, I., Webb, N. A., Wojtowicz, D. T., \& Leeuwen, J. van 2020, \aap, 640, A124
\bibitem[Revnivtsev et al.(2006)]{Revnivtsev06}
  Revnivtsev, M. G., Sazonov, S., Gilfanov, M., Churazov, E., \& Sunyaev, R. 2006, \aap, 452, 1, 169
\bibitem[Revnivtsev, Vikhlinin, and Sazonov(2007)]{Revnivtsev07}
  Revnivtsev, M. G., Vikhlinin, A., \& Sazonov, S. 2007, \aap, 473, 857
\bibitem[Reynolds et al.(2024)]{Reynolds24}
  Reynolds, M., et al. 2024, ATel, 16481
\bibitem[Sakurai et al.(2012)]{Sakurai12}
  Sakurai, S., Yamada, S., Torii, S., Noda, H., Nakazawa, K., Makishima, K., \& Takahashi, H. 2012, \pasj, 66, 1
\bibitem[Sakurai et al.(2014)]{Sakurai14}
  Sakurai, S., et al. 2014, \pasj, 66, 1
\bibitem[Shaw et al.(2022)]{Shaw22}
  Shaw, A. W., et al. 2022, \mnras, 516, 1, 124-137
\bibitem[Shidatsu et al.(2011)]{Shidatsu11}
  Shidatsu, M., et al. 2011, \pasj, 63, SP3, S785-S801
\bibitem[Shimura and Takahara(1995)]{Shimura95}
  Shimura, T., \& Takahara, F. 1995, \apj, 445, 780
\bibitem[Takahashi et al.(2008)]{Takahashi08}
  Takahashi, H., et al. 2008, \pasj, 60, SP1, S69--S84
\bibitem[Tashiro et al.(2025)]{Tashiro25}
  Tashiro, M., et al. 2025, \pasj accepted
\bibitem[Tarter, Tucker, and Wallace(1969)]{Tarter69}
  Tarter, C. B., Tucker, W. H., \& Salpeter, E. E. 1969, \apj, 156, 943
\bibitem[Uchida et al.(2025)]{Uchida25}
  Uchida, H., et al. 2025, \pasj accepted, arXiv:2503.20180
\bibitem[Uchiyama et al.(2013)]{Uchiyama13}
  Uchiyama, H., Nobukawa, M., Tsuru, T. G., \& Koyama, K. 2013, \pasj, 65, 1, 19
\bibitem[Wang, Gotthelf, and Lang(2002)]{Wang02}
  Wang, D. Q., Gotthelf, E. V., \& Lang, C. C. 2002, \nat, 415, 148–150
\bibitem[Wang, Dong, and Lang(2006)]{Wang06}
  Wang, D. Q., Dong, H., \& Lang, C. 2006, \mnras, 371, 1, 38--54
\bibitem[Wenger et al.(2000)]{Wenger00}
  Wenger, M., et al. 2000, \aaps, 143, 9
\bibitem[Yamauchi et al.(2016)]{Yamauchi16}
   Yamauchi, S., Nobukawa, K. K., Nobukawa, M., Uchiyama, H., \& Koyama, K. 2016, \pasj, 68, 4
\bibitem[Yuasa, Makishima, and Nakazawa(2012)]{Yuasa12}
   Yuasa, T., Makishima, K., \& Nakazawa, K. 2012, \apj, 753, 2, 129
\end{thebibliography}

\end{document}